\documentclass[aps,prapplied,reprint]{revtex4-1}
\usepackage[left=2cm, right=2cm, top=3cm, bottom=3cm]{geometry}
\usepackage{amsmath}
\usepackage{amsopn}
\usepackage{amsfonts}
\usepackage{natbib,graphicx,url,times,layouts,xcolor}
\usepackage{bm}
\usepackage{txfonts} 
\usepackage{cases} 
\usepackage{placeins} 
\definecolor{blueviolet}{rgb}{0.2, 0.2, 0.6}
\usepackage[pdftex, 
bookmarks=false, 
colorlinks=true,
allcolors=blueviolet,
pdfstartview={FitH}, 
]{hyperref}

\newcommand{\unit}{\ \mathrm}

\begin{document}

\title{Determining the position of a single spin relative to a metallic nanowire}

\author{J. F. da Silva Barbosa}
\affiliation{Universit\'e Paris-Saclay, CEA, CNRS, SPEC, 91191 Gif-sur-Yvette Cedex, France}

\author{M. Lee}
\affiliation{Department of Electrical Engineering, Pohang University of Science and Technology (POSTECH), 37673, Pohang, Korea}

\author{P. Campagne-Ibarcq}
\affiliation{Quantic Team, Inria Paris, 2 rue Simone Iff, 75012 Paris, France}

\author{P. Jamonneau}
\affiliation{Laboratoire Aim\'e Cotton, CNRS, Universit\'e Paris-Sud and Ecole Normale Sup\'erieure de Cachan, 91405 Orsay, France}

\author{Y. Kubo}
\affiliation{Okinawa Institute of Science and Technology Graduate University, Onna, Okinawa 904-0495, Japan}

\author{S.Pezzagna}
\affiliation{Department  of  Nuclear  Solid  State  Physics, Leipzig University, Germany}

\author{J. Meijer}
\affiliation{Department  of  Nuclear  Solid  State  Physics, Leipzig University, Germany}

\author{T. Teraji}
\affiliation{National Institute for Materials Science, 1-1 Namiki, Tsukuba, Ibaraki 305-0044, Japan}

\author{D. Vion}
\affiliation{Universit\'e Paris-Saclay, CEA, CNRS, SPEC, 91191 Gif-sur-Yvette Cedex, France}

\author{D. Esteve}
\affiliation{Universit\'e Paris-Saclay, CEA, CNRS, SPEC, 91191 Gif-sur-Yvette Cedex, France}

\author{R. W. Heeres}
\affiliation{Universit\'e Paris-Saclay, CEA, CNRS, SPEC, 91191 Gif-sur-Yvette Cedex, France}

\author{P. Bertet}
\affiliation{Universit\'e Paris-Saclay, CEA, CNRS, SPEC, 91191 Gif-sur-Yvette Cedex, France}

\email{patrice.bertet@cea.fr}


\begin{abstract}

The nanoscale localization of individual paramagnetic defects near an electrical circuit is an important step for realizing hybrid quantum devices with strong spin-microwave photon coupling. Here, we demonstrate the fabrication of an array of individual NV centers in diamond near a metallic nanowire deposited on top of the substrate. We determine the relative position of each NV center with $\sim$10\,nm accuracy, using it as a vector magnetometer to measure the field generated by passing a dc current through the wire. 

\end{abstract}

\maketitle


Single quantum defects have found a wide range of applications in quantum technologies~\cite{morton_hybrid_2011,degen_quantum_2017,awschalom_quantum_2018}. 
The Nitrogen Vacancy (NV) center in diamond is a defect with a ground-state electronic spin triplet that can be read-out optically at room-temperature at the single-defect level ~\cite{gruber_scanning_1997}. Because of these exceptional properties, NV centers can be used for single-spin magnetometry~\cite{taylor_high-sensitivity_2008}, enabling magnetic resonance spectroscopy~\cite{staudacher_nuclear_2013,aslam_nanoscale_2017} and probing of magnetic structures~\cite{tetienne_nanoscale_2014,thiel_quantitative_2016,gross_real-space_2017,thiel_probing_2019} at the nanoscale. 

NV centers are also promising for the implementation of quantum networks~\cite{togan_quantum_2010,bernien_heralded_2013,kalb_entanglement_2017,wehner_quantum_2018} at optical~\cite{faraon_resonant_2011,hausmann_coupling_2013} or microwave frequencies. In the latter case, the NV electronic spin is coupled to the magnetic field generated by a superconducting microwave resonant circuit deposited on top of the diamond substrate~\cite{kubo_strong_2010,kubo_hybrid_2011,zhu_coherent_2011}. Maximizing the spin-microwave coupling strength is essential for fast spin detection and entanglement generation. It was proposed to achieve that by inserting a narrow superconducting wire in the circuit and bringing the NV as close as possible to this nanowire~\cite{marcos_coupling_2010,haikka_proposal_2017}. This requires careful alignment of the nanowire on top of individual, shallow-implanted NVs. 


Precise positioning of NV centers relative to a nanostructure can be achieved by implanting nitrogen ions through masks of various types (pierced AFM tip~\cite{meijer_towards_2008,pezzagna_nanoscale_2010}, resist masks~\cite{toyli_chip-scale_2010}, metallic masks~\cite{scarabelli_nanoscale_2016}), or by direct implantation with a Focused-Ion-Beam~\cite{lesik_maskless_2013}. Here, we use Nitrogen implantation through nanometric holes pierced in a resist mask combined with electron-beam lithography alignment on etched marks in diamond. We fabricate an array of individual NVs with metallic nanowires on top of them, deposited on the diamond surface. 

\begin{figure}[!t]
  \includegraphics[width=3.3 in]{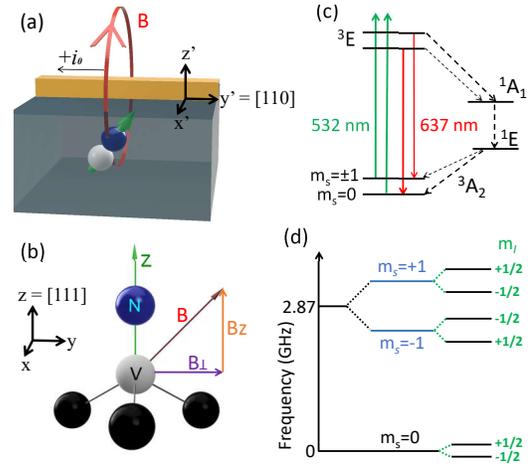}
  \caption{\label{Fig1}
  (a) Sketch of the experiment. 
  A NV center interacts with the magnetic field $B$ generated by the DC current $i_0$ passing through the nanowire. 
  The laboratory coordinate is indicated as $x'y'z'$ axes. The wire is directed along the $[110]$ crystalline direction.
  (b) Atomic configuration of a NV center. 
  The central gray ball indicates the vacancy, the blue ball the $^{15}$N atom, and the three black balls the carbon atoms.
  NV symmetry axis is denoted as $z$. $xyz$ is a coordinate system centered on the NV, with $z$ the NV axis along the $[111]$ crystalline axis.
  (c) Energy spectrum, showing the spin triplet ground and excited states, and two intermediate spin singlet states $^1A_1$ and $^1E$. For readout, the NV is excited by a green laser at $532\,\mathrm{nm}$, and the resulting photoluminescence (PL) in the red is detected. The NV can also relax non-radiatively, via the spin singlet states (dashed lines).
  (d) Schematic of the energy levels of the spin-triplet ($S=1$) ground state in a small magnetic field along $z$, showing the zero-field splitting $D/2\pi = 2.87$\,GHz between $m_S=0$ and $m_S=\pm 1$, as well as its hyperfine interaction with the $I=1/2$ $^{15}N$ atom. 
  }
\label{fig:setup}
\end{figure}

To characterize a posteriori the position of each NV relative to the nanowire, super-resolution optical methods such as STED~\cite{rittweger_sted_2009} or STORM cannot be used, as they would either melt the metallic wire or be blinded by the wire optical response. Here, we use single-NV vector magnetometry~\cite{chen_vector_2013} to measure the vector magnetic field generated when passing a dc current through the nanowire (see Fig.\ref{fig:setup}a), and infer the relative position of the NV with respect to the nanowire with a $\sim 10$\,nm precision. These measurements also yield the spin-microwave coupling strength achieved if the nanowire was inserted into a superconducting resonator~\cite{haikka_proposal_2017}.

The principle of single NV vector magnetometry~\cite{chen_vector_2013} relies on the sensitivity of the energy levels of an NV formed with a $^{15}N$ isotope atom (see Fig.\ref{fig:setup}b-d) to an externally applied magnetic field $\mathbf{B}$. This isotope is not naturally abundant, but $^{15}$NVs can be obtained through ion implantation~\cite{rabeau_implantation_2006}. The $S=1$ electron spin ground state is coupled by the hyperfine interaction to the nuclear spin $I=1/2$ of the $^{15}N$ nitrogen atom. The spin Hamiltonian of a NV center having its axis along the $z$ direction can be written as
\begin{equation}
    H = D S_z^2 + \gamma_e \mathbf{B} \cdot \mathbf{S} + \gamma_I \mathbf{B} \cdot \mathbf{I} + A_{z} S_z I_z + A_\perp (S_+ I_- + S_- I_+),
\end{equation}

\noindent with $D/2\pi = 2.87$\,GHz the electron-spin zero-field splitting, $\gamma_e/2\pi = 28$\,GHz/T the electron spin gyromagnetic ratio, $\gamma_I/2\pi = - 4.3$\,MHz/T the gyromagnetic ratio of the $^{15}N$ nuclear spin, $A_{z}/2\pi=3.03\unit{MHz}$ the secular component of the hyperfine interaction and $A_{\perp}/2\pi =3.65\unit{MHz}$ the non-secular component.


When a field $B_z$ is applied parallel to the NV axis $z$, the energy eigenstates are (in the secular approximation) $|m_S,m_I\rangle$ with $m_S=-1,0,+1$ ($m_I = \pm 1/2$) the eigenvalue of $S_z$ ($I_z$), and the transition frequencies between $|0,m_I\rangle $ and $|\pm 1,m_I \rangle$ read $\omega_{\pm,m_I} = D \pm \gamma_e B_z \pm m_I A_z $. When the field has in addition a small transverse component $|B_\perp| \ll D/\gamma_e$, both the frequency $\omega_{\pm,m_I}$ and the electron spin eigenstates are negligibly affected~\cite{chen_vector_2013}. Therefore, the standard measurement of $\omega_{\pm,m_I}$ using the spin-dependent optical transitions shown in Fig.\ref{fig:setup}c allows to infer only $B_z$~\cite{taylor_high-sensitivity_2008,maletinsky_robust_2012,rondin_nanoscale_2012,rondin_magnetometry_2014}.

Information about $B_\perp$ can nevertheless be obtained from the properties and dynamics of the $^{15}N$ nuclear spin, as described in ~\cite{chen_vector_2013}. Consider the nuclear spin Hamiltonian conditioned on the electron spin state. When $m_S=\pm 1$, it is dominated by $\pm A_z I_z$ (as long as $B_{\perp} \ll A_z / \gamma_I$, which we assume here), and the nuclear spin energy eigenstates are therefore $|m_I\rangle$. When $m_S=0$ however, only the nuclear Zeeman Hamiltonian remains; as a result, the states $|m_I \rangle$ become coupled by the term $\gamma_{I,\perp} B_\perp$, which gives access to a direct measurement of $B_\perp$. Note that the non-secular terms of the hyperfine interaction~\cite{childress_coherent_2006,chen_vector_2013} renormalize the Zeeman interaction, leading to an effective gyromagnetic ratio for the perpendicular component of the field $\gamma_{I,\perp} /2\pi = 75$\,MHz/T much larger than $\gamma_I$.



\begin{figure}[!t]
  \includegraphics[width=3.5 in]{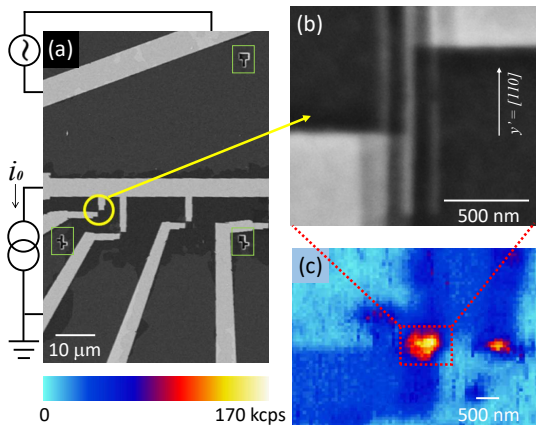}
  \caption{\label{Fig2}
  Sample micrographs. 
  (a) Scanning electron microscope image of the aluminium wires (white) fabricated on diamond (dark).
  The dc and microwave circuitry are drawn in black. Alignment marks are visible and pointed by green squares. 
  (b) Micrograph of the nanowire. Left and right shadows are due to the three- angle evaporation technique used.
  (c) Confocal microscope image of the photoluminescence of the aluminium and the NV center (red) behind the nanowire. 
  }
  \label{fig:image}
\end{figure}

Sample fabrication starts by the creation of an array of NV centers by ion implantation in a commercial electronic-grade, chemical-vapor-deposition-grown diamond chip (supplied by the Element 6 company). Alignment marks are first patterned by electron-beam lithography and etched into the diamond. An implantation mask is fabricated by opening an array of holes with $\sim 20$\,nm diameter in a 120-nm-thick polymethylmethacrylate (PMMA) resist layer using electron-beam lithography, at a well-defined position with respect to the marks. The sample is implanted by a beam of $^{15}$N$^{2+}$ nitrogen ions with a flux $\sim 2500\, \mathrm{N}/\mu \mathrm{m}^2$ at $7.5\,\mathrm{keV}$, which should lead to an implantation depth of $11 \pm 5$\,nm according to Stopping and Range of Ions in Matter (SRIM) simulations. It is annealed at 900$^{\circ}$C for 1 hour in vacuum to create the NV centers. It is afterwards cleaned in several steps: first, in a boiling 3:4:1 acid mixture of HNO$_3$:H$_2$SO$_4$:HClO$_4$ for 6 hours, then in a 3:1 mixture of H$_2$SO$_4$ and H$_2$O$_2$ at 120$^\circ$C for 2 hours (Piranha clean), and finally with an oxygen plasma. The sample is then optically characterized in a home-built confocal microscope~\cite{jamonneau_competition_2016}, which enables us to measure the Optically-Detected Magnetic Resonance (ODMR) spectrum using a $532$\,nm laser to polarize and readout the spin~\cite{jelezko_observation_2004}. Because of a low dose and N to NV conversion yield ($\sim 3\%$), only a few NV centers are observed at the implantation holes location. By measuring their ODMR spectrum, we can identify NVs with the $^{15}\mathrm{N}$ isotope, and thus confirm that they are implanted~\cite{rabeau_implantation_2006}. Five implanted NV centers having their axis along the $[111]$ crystalline direction are selected for the experiment. 

Aluminum electrodes and nanowires are then fabricated on top of the diamond by  electron-beam lithography followed by aluminum evaporation and liftoff, using the alignment marks to position one nanowire on top of each pre-selected NVs. Using three-angle evaporation through a suspended germanium mask~\cite{dolan_offset_1977}, we obtain a 20\,nm-thick nanowire (5\,nm titanium/15\,nm aluminum), connected to thicker (50\,nm) pads suitable for bonding. The nanowires have a length of $500$\,nm and typical width of $40$\,nm, as seen in Fig.~\ref{fig:image}b, and their direction is along $[110]$, corresponding to the situation depicted in Fig.~\ref{fig:setup}a. Each nanowire is connected to a common ground on one side, and to a separate pad on the other side, so that the current $i_0$ through each wire is set independently. A microwave antenna is also patterned in proximity of the nanowires to drive the NV electron spins. The pads are bonded to a printed-circuit board, and then connected to a DC current source and a microwave generator.



\begin{figure}[!t]
  \includegraphics[scale = 0.65]{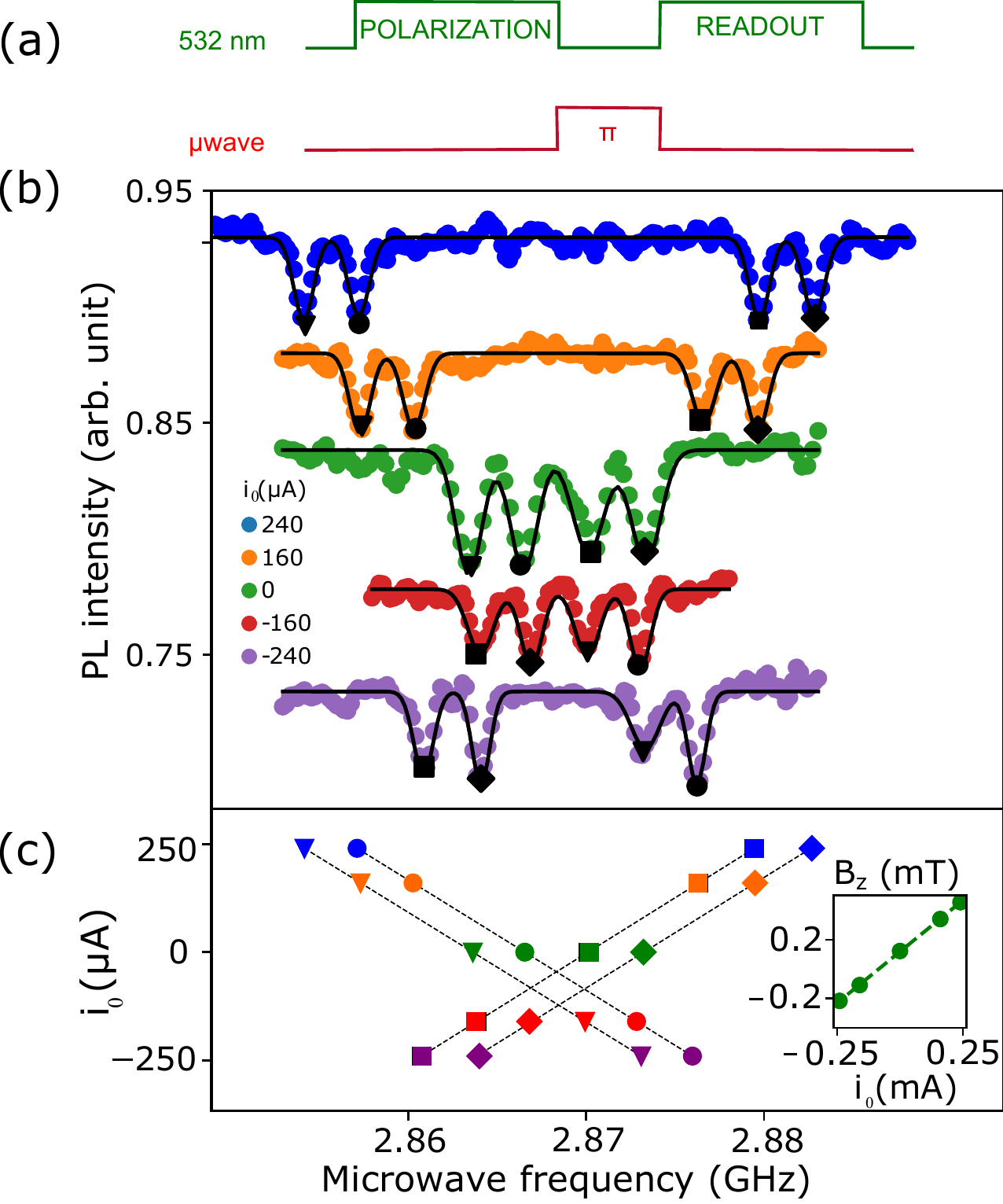}
  \centering
  \caption{\label{Fig3}
  (a) Laser (green) and microwave (red) pulse sequence of $i_0$-induced Zeeman shift measurement. (b) Measured spectrum of electron spin transitions for $i_0=$ 240 (blue), 160 (orange), 0 (green), -160 (red), and -240 (purple)~$\mu$A. 
  The black lines are fits to a sum of four Gaussian functions. Transitions from $|m_{s}=0\rangle$ to $|-1, -1/2\rangle$ (triangle), $|-1, +1/2\rangle$ (circle), $|+1, -1/2\rangle$ (square) and $|+1, +1/2\rangle$ (diamond). 
  (c) Plot of the centers of the Gaussian functions for 5 values of $i_0$. Dashed lines are linear fits. Inset: extracted value of $B_z$ as a function of $i_0$ (green dots), and linear fit (dashed line) yielding $\alpha_z = 1.4 \pm 0.1 \mathrm{T/A}$. 
  }
\label{fig:zeeman}
\end{figure}

\begin{figure*}[!t]
\includegraphics[scale = 0.45]{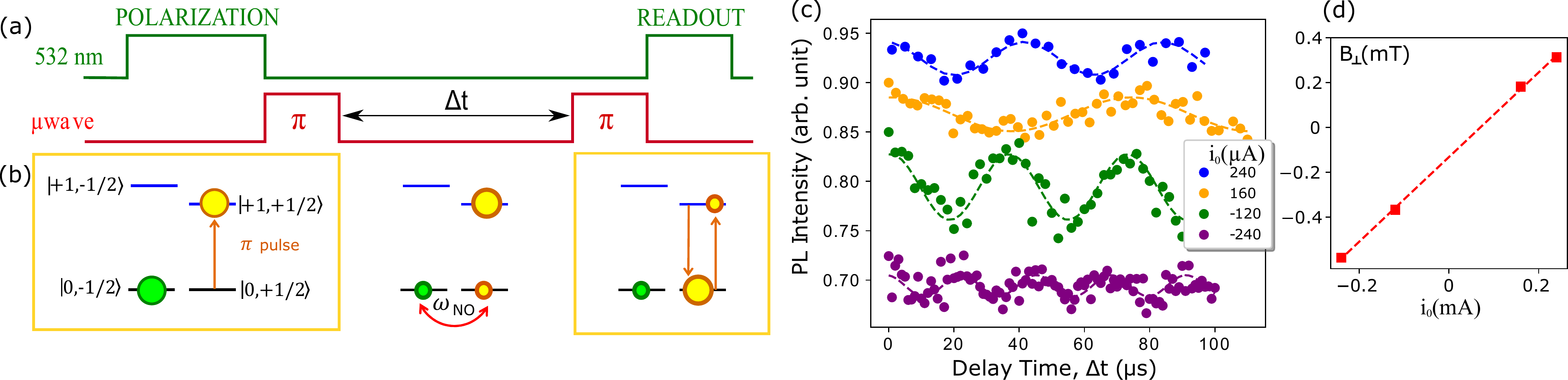}
  \centering
  \caption{\label{Fig4}
  (a) Laser (green) and microwave (red) pulse sequence, with $\Delta t$ the delay time between two $\pi$ pulses.
  (b) Sketched time evolution of spin state populations during the sequence.
  (c) Oscillation of the nuclear spin population measured through the PL intensity variation versus $\Delta t$, for different values of the dc current $i_0= 240$ (blue), 160 (orange), -120 (green), -240 (purple)~$\mu$A.
  The dashed lines are sinusoidal functions fitted to the data.
  (d) Calculated $B_{\perp}$ component (squares), and linear fit (dashed line) yielding $\alpha_\perp = 1.9 \pm 0.3 \mathrm{T}/\mathrm{A}$.
  }
\label{fig:nuclear_spin}
\end{figure*}

Because the nanowire width is much smaller than the wavelength, the NV photoluminescence (PL) can be detected despite the aluminum on the surface, provided the wire is long enough (i.e., longer than the spot size in the confocal setup) to avoid shadowing by the pads. This is visible in Fig.~\ref{fig:image}c, which shows a confocal scan around one of the nanowires. Some amount of photoluminescence is measured on top of the aluminum pads, but a bright spot is observed at the expected nanowire and NV location. The count rate is higher than the typical NV count rate ($\sim 100$kcps), probably because of stray photoluminescence from the wire or from dirt around it. Out of the $5$ selected NVs, two showed no ODMR signal, either due this stray photoluminescence or to a problem of charge instability~\cite{bluvstein_identifying_2019}. For the other $3$ NVs, an ODMR contrast between $2$ and $5\%$ was observed. The spectrum is shown in Fig.~\ref{fig:zeeman}b for different values of $i_0$ for one NV. Two doublet peaks are observed, corresponding to the $\omega_{\pm,m_I}$ frequencies~\cite{rabeau_implantation_2006}. The linear dependence of $\omega_{\pm,m_I}(i_0)$ shown in Fig.~\ref{fig:zeeman}c yields $\alpha_z \equiv dB_z / di_0 = 1.4 \pm 0.1 \mathrm{T}/\mathrm{A}$, the finite $B_z (0)$ being the earth magnetic field. 


To measure the orthogonal field, the pulse sequence shown in Fig.~\ref{fig:nuclear_spin} is applied. After initialization of the electron spin in $m_S=0$ by a green laser pulse, the NV is found with equal probability in states $|0,+1/2\rangle$ and $|0,-1/2\rangle$. A nuclear-spin-state-selective $\pi$ pulse is then applied to the electron spin transition. This pulse transfers all population from $|0,+1/2\rangle$ into $|+1,+1/2\rangle$. Whereas the latter is an energy eigenstate (and therefore does not evolve in time), $|0,-1/2\rangle$ is coupled to $|0,+1/2\rangle$ by $\gamma_{I,\perp} B_\perp$, leading to an oscillation with a frequency $\omega_{NO} = \sqrt{(\gamma_{I}B_{z})^2 + (\gamma_{I,\perp}B_{\perp})^2 }$ between the two states. This oscillation is detected by letting the system evolve for a variable delay $\Delta t$, followed by another nuclear spin selective $\pi$ pulse on the same transition which maps the nuclear-spin-oscillation into an oscillation between $m_S=0$ and $m_S=+1$ detectable by a readout laser pulse following the $\pi$ pulse. 

Data are shown in Fig.~\ref{fig:nuclear_spin}c, for a range of currents, on the same NV center. Oscillations in the photoluminescence as a function of $\Delta t$ are observed over a range of $100 \mu \mathrm{s}$, without measurable decay, as expected for a $^{15}\mathrm{N}$ nuclear spin whose coherence time should be limited by the electron spin energy relaxation time, on the order of one millisecond. The oscillation frequency is seen to depend linearly on $i_0$, allowing us to extract the value of $\alpha_\perp \equiv dB_{\perp}/di_0 = 1.9 \pm 0.3 \mathrm{T}/\mathrm{A}$. 

\begin{figure*}[!t]
  \includegraphics[width=6.8 in]{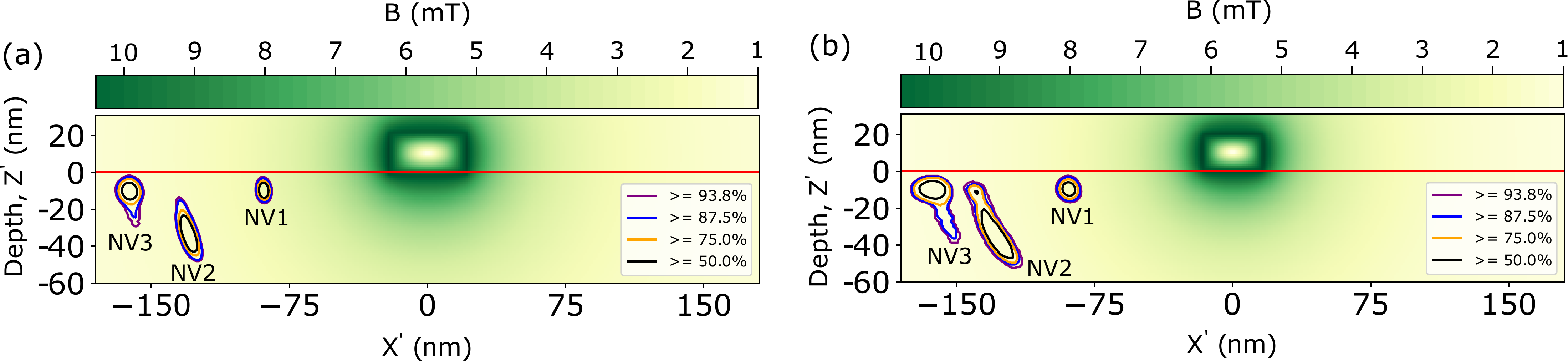}
  \caption{\label{Fig06_01.pdf}
  Estimating single NV centers positions.
  The color map represents the magnitude of the magnetic field $B$ due to a DC current of $1$\,mA passing through the nanowire. 
  The continuous red line at depth 0\,nm is the diamond surface.
  The colored lines show iso-probability contours of the NV location taking into account the uncertainties on the nanowire geometry and the measured field spatial derivatives (see text). (a) Infinitely long nanowire.
  (b) Finite-length nanowire. 
  }
\label{fig:NV_position}
\end{figure*}

We then determine the position of the measured NVs by considering the magnetic field generated by an ideal infinite wire of thickness $t$ and width $w$ carrying a current $i_0$. By knowledge of the NV axis orientation, we can uniquely determine the position $(x',z')$ which most closely matches the measured magnetic field-to-current-ratios $(\alpha_z,\alpha_{\perp})$. We proceed to determine a probability density function (pdf) for the NV position using a bootstrapping method: we perform the fitting procedure 5000 times using Gaussian-distributed parameters, to account for the uncertainty in the wire width $w=36 \pm 5 \,\mathrm{nm}$, thickness $t=20 \pm 2 \,\mathrm{nm}$; we also account for an uncertainty $\sigma = 2 \cdot 10^{-2}\, \mathrm{mT/mA}$ in the determined $\alpha_z$ and $\alpha_\perp$. This results in the position pdf shown in Fig.~\ref{fig:NV_position}a. To take into account the fact that our nanowire is not an ideal infinite wire, we perform finite element method simulations to determine the magnetic field deviation if the NV center is not positioned exactly in the center along the long axis. We find a relative uncertainty for $B_{\perp'}$ of $1.1\%$ and for $B_{z'}$ of $3.7 \%$. Using again the fitting procedure described before, but with additional errors on the generated magnetic fields, we find slightly larger error bars. The corresponding pdf's are shown for the three NVs in Fig.\ref{fig:NV_position}b, and all positions are summarized in Table 1.


\begin{table}[h]
    \centering
    \begin{tabular}{c|c c|c c}
         & $x_1'$ (nm) & $z'_1$ (nm) & $x_2'$ (nm) & $z_2'$ (nm)  \\
         \hline
         NV1 & $-83.9 \pm 0.8$ & $-8.6\pm 2.7$ & $-83.9 \pm 2.4$ & $-8.6 \pm 2.7$ \\
         NV2 & $-122.6 \pm 2.9$ & $-30.1 \pm 7.5$ & $-122.7 \pm 6.7$ & $-29.2 \pm 11.2$ \\
         NV3 & $-152.3 \pm 2.6$ & $-11.1 \pm 6.2$ & $-151.9 \pm 5.7$ & $-13.4 \pm 9.2$ \\
         \hline
    \end{tabular}
    \caption{Position of the NV centers relative to the nanowire, assuming an infinitely-long nanowire (left, $[x'_1,z'_1]$) or taking into account finite-length corrections as explained in the text (right, $[x'_2,z'_2]$).}
    \label{tab:positions}
\end{table}

As an independent check of the validity of our estimate, we note that the depth determined for NV1 and NV3 is in qualitative agreement with the $11 \pm 5\,\mathrm{nm}$ depth expected from SRIM simulations given the implantation energy of $7.5\, \mathrm{keV}$, while the larger depth of NV2 may be due to channeling. Further confirmation could be brought by measuring the depth using the proton signal from the diamond surface\cite{pham_nmr_2016}, although this was not done in the present experiment. The average lateral position of the three NVs is shifted by $\sim 120\mathrm{nm}$ from the nanowire center, which may be due to a systematic shift during the two alignment steps of the electron-beam writing. The standard deviation in lateral positioning of $27\,\mathrm{nm}$ can be semi-quantitatively understood as arising from the finite size of the implantation aperture (with $\sim 20 \, \mathrm{nm}$ diameter) and the straggle upon implantation (also $\sim 20 \, \mathrm{nm}$ diameter).

An application of our method is to determine, solely from room-temperature measurements, the coupling constant $g$ of a NV to a nanowire resonator, as proposed in~\cite{haikka_proposal_2017} for its detection using a circuit-QED architecture at millikelvin temperatures. Indeed, this determination simply requires the knowledge of the vacuum fluctuations $\delta B_1$ of the magnetic field component of the resonator mode at the NV location, since $g=\gamma_e \delta B_{1,\perp} \langle m_S = 0 | S_x | m_S = -1 \rangle$ (with $\langle m_S = 0 | S_x | m_S = -1 \rangle = 1/\sqrt{2}$ for a spin 1). Consider for instance the resonator design envisioned in~\cite{haikka_proposal_2017}, for which quantum fluctuations of the current $\delta i =35$\,nA were estimated by finite-element simulations. The resulting $\delta B_{1,\perp} = \alpha_{\perp} \delta i$ is directly obtained for each of the $3$ NVs measured, yielding a coupling constant $g/2\pi=0.6$, $0.7$, and $1$\,kHz. We stress that in this coupling constant estimate, the uncertainty would arise mostly from the uncertain knowledge of $\delta i$, since $\alpha_{\perp}$ on the other hand is measured with a few percent precision. We can then estimate the single-spin detection time $\kappa^2 \gamma_2 / (\eta g^4)$~\cite{haikka_proposal_2017}, with $\kappa$ the resonator energy damping rate, $\gamma_2$ the NV coherence time, and $\eta$ the microwave detection efficiency. Taking $\eta=1$, $\kappa = 10^5\,s^{-1}$, $\gamma_2 = 10^5\,s^{-1}$, we predict a measurement time between $0.6$ and $5$\,s for unit signal-to-noise ratio. Reducing the systematic alignment shift by better alignment procedure in the electron-beam lithography appears therefore necessary for single-spin microwave detection within millisecond integration times, as proposed in \cite{haikka_proposal_2017}.





We have measured the position of individual nitrogen-vacancy (NV) centers with respect to a metallic nanowire on diamond, using single-NV vector magnetometry~\cite{chen_vector_2013}, with a precision of $\sim 10\,\mathrm{nm}$. The lateral positioning of the implanted NVs shows a systematic shift of $\sim 120 \,\mathrm{nm}$ possibly due to misalignment during e-beam lithography. The spread of lateral position is compatible with the implantation mask dimensions and expected straggling. Our method enables in particular a direct room-temperature measurement of the coupling constant of individual spins to a nanowire microwave resonator.



\emph{Acknowledgements} We acknowledge discussions with M. Pioro-Ladri\`ere, as well as technical support from P.~S\'enat and P. F. Orfila, D.~Duet, J.-C.Tack, A. Forget. We acknowledge support of the European Research Council under the European Community’s Seventh Framework Programme (FP7/2007–2013) through grant agreement No.~615767 (CIRQUSS), and of the Chaire Industrielle NASNIQ  under contract  ANR-17-CHIN-0001 cofunded by Atos. T.T. acknowledges the support of JSPS KAKENHI (No. 20H02187, 19H02617 and 16H06326), JST CREST (JPMJCR1773) and MEXT Q-LEAP (JPMXS0118068379).

\emph{Data availability} The data that support the findings of this study are available from the corresponding author upon reasonable request.



%

\end{document}